\documentclass[11pt,preprint]{emulateapj}

\usepackage{natbib}

\shorttitle{Searching for ICM dust}
\shortauthors{Muller et al. 2008}

\begin{document}

\title{Searching for dust in the intracluster medium from reddening of background galaxies}

\author{Muller, S.\altaffilmark{1}, Wu, S.-Y.\altaffilmark{1,2}, Hsieh, B.-C.\altaffilmark{1},
Gonz\'alez, R. A.\altaffilmark{3},
Loinard, L.\altaffilmark{3},
Yee, H. K. C.\altaffilmark{4}
\& Gladders, M. D.\altaffilmark{5}}

\affil{\altaffilmark{1} Academia Sinica Institute of Astronomy and Astrophysics, P.O. Box 23-141, Taipei, 106 Taiwan}
\affil{\altaffilmark{2} Current address: Department of Astronomy, University of Massachusetts, Amherst, MA 01106, USA}
\affil{\altaffilmark{3} Centro de Radioastronom\'ia y Astrof\'isica,
%Instituto de Astronom\'ia, Universidad Nacional Aut\'onoma de M\'exico, 
58190 Morelia, Michoac\'an, Mexico}
\affil{\altaffilmark{4} Department of Astronomy and Astrophysics, University of Toronto, ON, M5S 3H4, Canada}
\affil{\altaffilmark{5} Department of Astronomy and Astrophysics, University of Chicago, Chicago, IL, USA}

\begin{abstract}

We report a search for the presence of dust in the intra-cluster medium based on the study of statistical
reddening of background galaxies. Armed with the Red Sequence Cluster survey data, from which we extracted
({\em i}) a catalog of 458 clusters with $z_{clust} < 0.5$ and
({\em ii}) a catalog of $\sim$ 90,000 galaxies with photometric redshift $0.5 < z_{ph} < 0.8$ and photometric
redshift uncertainty $\delta z_{ph} / (1+z_{ph}) < 0.06$, we have constructed several samples of galaxies 
according to their projected distances to the cluster centers.
No significant color differences [$<$E($B-R_c$)$>$ = $0.005 \pm 0.008$, and $<$E($V-z'$)$>$ = $0.000 \pm 0.008$] 
were found for galaxies background to the clusters, compared to the references.
Assuming a Galactic extinction law, we derive an average visual extinction of $<$A$_V$$>$ = $0.004 \pm 0.010$
towards the inner $1\times R_{200}$ of clusters.

\end{abstract}


\section{Introduction}

\cite{zwi51} was the first one to observe a weak and diffuse intra-cluster light in the Coma cluster, suggesting
the presence of material in the intra-cluster medium (ICM). The jump in sensitivity of the CCD detectors later
allowed intra-cluster light to be studied more quantitatively (see e.g., \citealt{ber95,fel02,zib05}). The
existence of an intra-cluster stellar population has further been clearly established with the detection of
planetary nebulae \citep{arn96,fel98}, red giant stars \citep{fer98} and supernovae \citep{gal03} in the ICM.
In parallel, X-ray spectroscopic observations revealed the presence of heavy metals in the ICM \citep{mit76,mus96}.
This ICM matter is commonly believed to originate from stripping and/or disruption of cluster galaxies, and
one can then naturally wonder about the existence of dust in the ICM.

It is true that the ICM is a harsh medium, where strong X-ray radiation and gas temperatures of $\sim$ 10$^7$ K
may impose severe restrictions on the presence of dust. However, the survival timescale for dust grains
in this environment is still of the order of few 10$^8$ yr \citep{dra79}. Besides, many mechanisms, such as tidal
or ram pressure stripping, galactic winds, mass loss from ICM red giants and supergiants, or possible accretion
of primordial dust, may contribute to the release or replenishment of dust in the ICM [see, e.g.,\cite{pop00},
\cite{sch05}, \cite{dom06} for predictions on the efficiency of the different mechanisms].

While there have been many attempts, using various methods, to detect the existence of dust in the ICM (see
Table~\ref{prevsearch}), the debate is still open. The methods are mostly twofold: either based on a direct
search for dust thermal emission, typically peaking in the far infrared to submillimeter range, or on an
indirect search for extinction and reddening of background sources.

The direct detection of extended thermal emission in the intra-cluster medium would constitute the best evidence
for the presence of ICM dust. Theoretical predictions of the IR emission
have been made (see e.g., \citealt{dwe90}, \citealt{yam05}); taking into account dust injection by galaxies,
sputtering, and heating by the hot ICM plasma, they yield a mean dust temperature of $\sim 20 - 30$ K, and
dust-to-gas ratio $\sim 10^{-6}$. \cite{ann93} could not detect significant emission from a submillimeter search
towards 11 clusters with cooling flows, up to a limit of $\sim 10^8$ M$_\odot$. \cite{wis93} observed with IRAS a
larger sample of 56 clusters, from which only two may show evidence for extended diffuse emission. Further ISO
observations revealed  FIR excess emission towards Coma \citep{sti98}, possibly due to a dust mass of $\sim
10^7 - 10^9$ M$_\odot$ in the inner 0.4 Mpc region, but none towards five other Abell clusters \citep{sti02}.
More recently, \cite{mon05} adopted a statistical approach and co-added IRAS maps of a large number ($> 11,000$)
of clusters and detected a significant emission in all four IRAS band (12, 25, 60 and 100 $\mu$m). The origin of
the emission, tentatively attributed to ICM dust, is, however, questionable. Indeed, in addition to the
challenging detection of low surface brightness and extended emission, IR observations of ICM dust might be
affected by contamination from Galactic cirrus (e.g., \citealt{wis93,bai07}), as well as from cluster members
\citep{qui99}, dusty star forming galaxies in clusters (\citealt{gea06}), dust associated with the central
dominant galaxies or the presence of a cooling flow \citep{bre90,gra90,cox95,edg99}, or even from background
(possibly magnified) galaxies.

The first claim for ICM dust was, nevertheless, historically made by \cite{zwi62}, based on the extinction of
light from distant clusters by nearby ones. He estimated A$_V \sim 0.4$ for Coma. \cite{kar69} derived a
similar value of A$_V \sim 0.3$ mag for Coma, and found a mean cluster extinction of A$_V \sim 0.20 \pm 0.05$
mag by averaging over 15 clusters. \cite{bog73} measured that distant Abell clusters appear to be inversely
correlated on the sky with nearby ones. To account for this effect, they argued for an extinction corresponding
to A$_{V} \sim 0.4$ mag and extending to $\sim 2.5$ times the optical radii of the nearby clusters. However,
these first results were probably severly affected by the difficulties of identifying background clusters.

Obvious background sources to search for dust extinction and reddening are quasars. \cite{boy88}
measured a value of A$_B = 0.2$ mag (A$_ V \sim 0.15$ mag) from a deficiency of ultraviolet-excess-selected
objects behind clusters of galaxies. Similarly, \cite{rom92} concluded that an area-averaged extinction of
A$_B \sim 0.5$ mag was required to explain the anti-correlation between quasars and Abell clusters. On the
other hand, \cite{mao95} could not measure a significant difference [E($B-V$)$\leq 0.05$] between the color
distribution of radio-selected quasars located within 1$\degr$ of a foreground Abell cluster, and quasars
with no cluster in their line of sight. He reached the conclusion that the apparent deficiency of optically
discovered quasars in the background of clusters could be attributed to selection effects, reflecting the
difficulty to identify quasars in crowded fields.

More recently, \cite{nol03} proposed, to our knowledge, the first and only search so far of ICM dust based
on extinction and reddening of background galaxies. They used a catalog of $\sim 140$ nearby ($z < 0.08$)
clusters with $B$ and $R$ photometric data, complete to $B \sim 20.5$ mag and $R \sim 19.5$ mag. For each
cluster, they constructed two samples. The ``cluster'' sample includes all galaxies lying in a circular
region centered on the cluster, and with an angular radius equivalent to the physical radius of 1.3 Mpc at
the distance of the cluster. The second sample, namely ``control sample'', contains galaxies projected
within a ring of inner radius 1.3 Mpc and of outer radius such that the area covered by the cluster and
control samples are identical. The cluster and control samples for all clusters were then respectively
merged to increase the number of galaxies and improve the statistics. The cluster sample inevitably suffers
from contamination of cluster members and foreground galaxies, although \cite{nol03} estimate that these
latest amount to less than 10\% contamination. The cluster and control group galaxies were then distributed
in two color-magnitude diagrams (CMD), and any extinction and reddening of the cluster group was searched
by shifting the cluster CMD both in magnitude and color with respect to the control CMD, until the two CMD
matched. The pixel resolution (in magnitude and color) of the CMD is constrained by the number of galaxies
in each pixel, which should be high enough. Given the number of galaxies in each of their samples ($<10^5$),
\cite{nol03} fixed the smallest reasonable pixel size for $\Delta$($B-R$) and $\Delta R$ to 0.025. They
could not detect a significant difference between the cluster and control samples, thus yielding upper
limits of A$_R < 0.025$ and E($B-R$)$< 0.025$ mag (A$_V < 0.044$) for ICM dust.  

We note briefly that dust extinction and reddening of background galaxies have been successfully detected
in galactic disks (see, e.g., \citealt{gon98}, \citealt{hol05}), galactic halos \citep{zar94}, and, recently,
in the intergalactic medium towards the M81 group \citep{xil06}. This simple method should become more and
more popular with the advent of wide field cameras on board of large telescopes.

Finally, in a very recent paper, \cite{che07} claim the detection of reddening of background quasars behind
$0.1 < z < 0.3$ galaxy clusters using the {\em Sloan Digital Sky Survey} (SDSS) photometric and
spectroscopic data. They report E($g-i$) of $0.003 \pm 0.002$ from photometry analysis, on one hand, and
$0.008 \pm 0.003$ from spectroscopic extinction curve, on the other hand, towards the central $\sim 1$ Mpc
of clusters. Assuming a Galactic extinction curve (\citealt{sch98}), this converts to A$_V \sim 0.005 \pm 0.003$
and $0.013 \pm 0.004$, respectively. The reference quasars were taken at large projected distances ($> 7$ Mpc)
from the cluster centers. Their measurement of average reddening at different radial annuli seems
to indicate both a large covering factor and a large scale distribution of dust, up to $5-6$ Mpc.

In the present paper, we propose to address the question of the presence of dust in the ICM by studying
the statistical reddening of galaxies located in the background of clusters, as compared to a reference
sample of galaxies located away from the line of sight to clusters. We have constructed the samples based
on the cluster and galaxy catalogs from the Red-sequence Cluster Survey (RCS, \citealt{gla05}), and using
galaxy photometric redshifts. In \S2, we briefly describe the RCS data and present our galaxy sample
selection and analysis. Our results are discussed in \S3, and a summary is given in \S4.

\section{Data and Data Analysis}

\subsection{The Red-Sequence Cluster Catalog} \label{RCS}

The Red-Sequence Cluster Survey (RCS, \citealt{gla05}) made use of both the Canada-France-Hawaii Telescope
(CFHT) and the Cerro Tololo Inter-American Observatory to cover a total sky area of $\sim$90 deg$^{2}$ in
the $R_c$ (6500\AA) and $z'$ (9100\AA) bands. We hereby consider only the CFHT RCS data overlapping with
the CFHT follow-up observations in the $B$ and $V$ bands presented in \cite{hsi05}, briefly described below
in \S\ref{photoz}. The intersection of the $B, V, R_c$ and $z'$ data results in a total of 33.6 deg$^{2}$
over 10 different fields (see Table~\ref{RCSsubset}), located at high Galactic latitude ($>$ 40$\degr$)
to minimize Galactic extinction.

The RCS was originally designed to obtain a large sample of galaxy clusters up to redshift $z \sim 1.4$,
using the cluster red-sequence method \citep{gla00}. In this method, clusters are identified as density
enhancements in the four-dimensional space of position, color and magnitude. Red-sequence models provide
cluster photometric redshifts for which a comparison with spectroscopic data shows that the accuracy is
typically better than 0.05 over the redshift range $0.2 < z < 1.0$.

In total, the RCS cluster catalog includes $\sim$ 4,000 identified clusters with redshift $0.2 < z < 1.4$,
each listed with position, redshift, size and richness [see \cite{gla05}]. The cluster size is defined by
the $R_{200}$ radius, within which the average galaxy density is 200 times the critical density. The cluster
richness is estimated through the amplitude of the cluster galaxy correlation function Bgc. This parameter
basically traces the excess galaxy counts around a reference point (e.g., the center of a cluster), given a
luminosity function and a spatial distribution for galaxies, and is known to be a robust quantitative
measurement of the cluster richness [see \cite{yee99} for a detailed discussion on the Bgc parameter]. For
this study, we extracted only clusters with $z_{clust}$ $< 0.5$ and Bgc $> 200$ (in unit of 
Mpc$^{1.8}$~$h_{50}^{-1.8}$, where $h_{50} = H_0 /50$~km~s$^{-1}$~Mpc$^{-1}$), yielding a subset catalog
of 458 clusters. The typical uncertainty in Bgc for RCS clusters is 150 to 200, and the average Bgg (i.e.,
the galaxy-galaxy correlation amplitude computed by picking any random galaxy) is about 60. 

The RCS cluster catalog is not complete for nearby clusters (i.e., $z < 0.2$), due to the choice of the $R_c$
and $z'$ filters, optimized to find clusters at higher redshift. We have used the MaxBCG cluster catalog issued
from the SDSS, recently published by \cite{koe07}, to check the apparent surface covered by SDSS nearby clusters
($0.05 < z < 0.35$), with respect to RCS data. 7 out of the 10 RCS patches overlap with the SDSS and we found a
total of 41 SDSS clusters within these limits. By fixing a typical size of 5$\arcmin$ (in radius) for each cluster,
we estimate that the coverage of these nearby clusters represents a surface area less than 4\% of the corresponding
RCS data. We therefore neglect the contamination by nearby ($z < 0.2$) clusters in our study.

\subsection{Photometric Redshifts} \label{photoz}

Additional photometric data were obtained in the $V$ and $B$ bands with the CFHT for some part of the RCS fields.
The intersection of $B, V, R_c$ and $z'$ data covers a total $\sim 33$ deg$^2$, with limiting magnitudes of 23.9
in $z'$ (AB), 25.0 in $R_c$ (Vega), 24.5 in $V$ (Vega), and 25.0 in $B$ (Vega). The four-band photometry data,
corrected for Galactic extinction, were used by \cite{hsi05} to derive photometric redshifts for $\sim$ 1.2 million
galaxies, using an empirical quadratic polynomial fitting technique with spectroscopic redshifts for 4,924 galaxies
in the training set. However, since then more spectroscopic data became available for the RCS fields, which can be
added to the training set. In particular, the DEEP2 DR2 \footnote{http://deep.berkeley.edu/DR2/} provides 4,297
matched spectroscopic redshifts in the range $0.7 < z < 1.5$. The new training set not only contains almost twice
as many objects compared to the original one but also provides a much larger sample for $z > 0.7$, i.e., better
photometric redshift solution for high redshift objects. Besides having a larger training set, we also used an
empirical third-order polynomial fitting technique with 16 Kd-Tree cells for the photometric redshift estimation,
allowing a higher redshift accuracy and giving a rms scatter $< 0.05$ for $\delta z_{ph}$ within the redshift
range $0.2 < z < 0.5$, and $< 0.09$ for galaxies at $0.0 < z < 1.2$. The refined calibration procedure and
photometric redshift method will be described in detail elsewhere (Hsieh et al., in prep.)

\subsection{Sample Selection and Analysis} \label{sampleselection}

The goal of our study is to measure any statistical color difference between a sample of galaxies seen in the
background of clusters and a reference sample of galaxies, presumably not affected by intervening dust. We first
extracted only the galaxies with $0.5 < z_{ph}$ $< 0.8$ (i.e, at redshift larger than that of selected clusters)
and $\delta z_{ph}/(1+z_{ph}) < 0.06$ (i.e., with good photometric redshift accuracy). A total of $\sim$ 90,000
galaxies satisfied these criteria. Each galaxy of our RCS subset is defined by its photometric redshift $z_{ph}$
and $B, V, R_c, z'$ magnitudes (with the associated uncertainties $\delta z_{ph}$, $\delta B, \delta V, \delta R_c$
and $\delta z'$).

Our samples were built by using a Monte Carlo method, in order to take into account the redshift and photometric
uncertainties. We repeated the process explained hereafter 100 times. Before checking its line of sight, we
replaced any given galaxy by a 'new' galaxy with:
\begin{equation}
z_{ph} \rightarrow z_{ph} + G(\delta z_{ph}),
\end{equation}
and
\begin{equation}
i-j \rightarrow (i + G(\delta i)) - (j + G(\delta j)),
\end{equation}
where i and j are any two filters taken from ($B, V, R_c, z'$), and G(x) is a random value following a Gaussian
distribution with standard deviation x. Two independent colors, $B-R_c$ and $V-z'$, were derived.

\begin{figure}
\includegraphics[width=8cm]{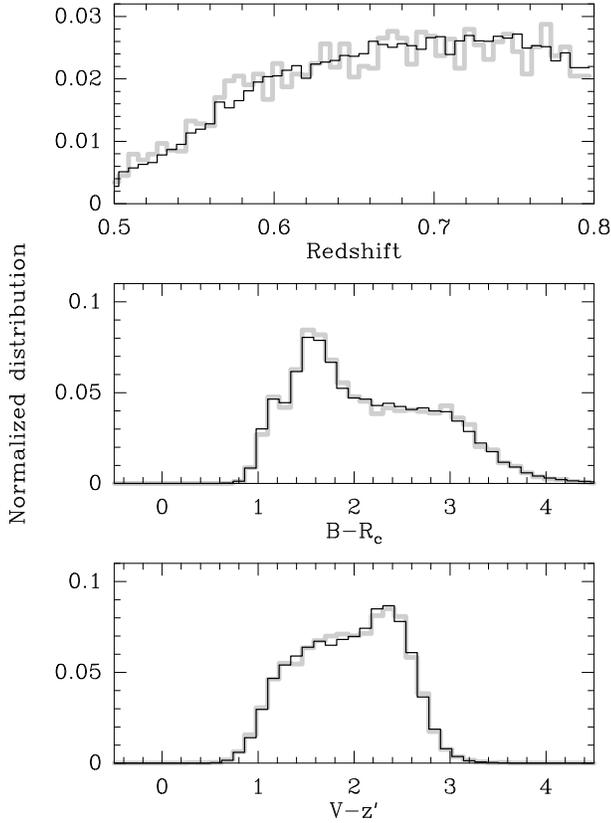}
\caption{Comparison of the redshift and color distributions of galaxies in sample $a$ ($< 1 \times$R$_{200}$;
{\em thick grey curve}) and sample $d$ ($[3-4]\times$R$_{200}$; {\em black curve}) for a given Monte Carlo
realization. All distributions were normalized to the total number of galaxies in each sample.}
\label{color}
\end{figure}

The line of sight to each galaxy was investigated by checking against the positions of all clusters in our
RCS cluster subset. We constructed five samples, containing all galaxies with redshifts such that
$z_{ph} - z_{clust} > 0.3$ and located at projected distances $l < 1\times R_{200}$ (sample $a$),
$l \in [1-2]\times R_{200}$ (sample $b$), $l \in [2-3]\times R_{200}$ (sample $c$), 
$l \in [3-4]\times R_{200}$ (sample $d$), and $l \in [6-7]\times R_{200}$ (sample $e$), 
respectively, from a cluster center.
A bootstrap technique (\citealt{efr79}) was applied to rebuild each sample, by randomly picking out N times
a galaxy amongst the N galaxies of the sample. Each 'new' sample was further binned in each color with a bin
resolution of $\delta$=0.05 magnitude. Fig.\ref{color} illustrates the redshift and color distributions obtained
for one given Monte-Carlo+bootstrap realization. We finally checked any color difference between two samples by
progressively shifting the color distribution of one relatively to the other by a bin offset $\Delta$
($\Delta$ being an integer), and calculating the value:
\begin{equation}
\chi (\Delta) = \sum_i ({\rm sample}1(i-\Delta)-{\rm sample}2(i))^2.
\end{equation}
The color difference $\Delta_{COL}$ between the two samples was then assumed to be the minimum value for
$\chi (\Delta)$ (i.e., that for which the two samples are most similar), calculated as:
\begin{equation}
\Delta_{COL} = \sum_\Delta \frac{\Delta \delta}{\chi^2(\Delta)} / \sum_\Delta \frac{1}{\chi^2(\Delta)}.
\end{equation}
In practice, we limited the shifts $\Delta$ between $-$10 to +10 bins, thus exploring a color difference
range $-$0.5 to +0.5 magnitude. For each Monte Carlo realization, we performed a total of 100 bootstrap runs,
which allowed us to derive a statistical dispersion associated with $\Delta_{COL}$.

The combination of Monte Carlo + bootstrap techniques allowed us (1) to take into account the redshift and
photometric uncertainties, and (2) to derive the average color differences between the samples, with their
associated dispersion. To give some numbers, we counted $\sim 8,000$ galaxies in sample $a$, $\sim 23,000$
in sample $b$, $\sim 36,000$ in sample $c$, $\sim 47,000$ in sample $d$ and  $\sim 73,000$ in sample $e$.
The final results for the color differences between the various samples are presented in Table~\ref{results}.

\section{Discussion}

Our results indicate that there is statistically no color difference between the samples of background galaxies seen
behind the inner part of clusters ($<1\times R_{200}$) and in the periphery up to $[6-7]\times R_{200}$. We
emphasize that while the color differences are derived from independent filter pairs with independent calibration,
the results in columns 2 and 3 of Table~\ref{results} are all consistent with no reddening, given the uncertainties.
Assuming a Galactic extinction law, we convert the E($B-R_c$) and E($V-z'$) values into visual extinction, A$_V$,
from the following relations: A$_V = 1.9 \times$E($B-R_c$) and  A$_V = 1.9 \times$ E($V-z'$) (taking the coefficients
from \citealt{sch98}), and calculate the weighted average of the two independent values (column 4 in Table~\ref{results}).
We end up with a visual extinction of $<$A$_V$$>$ = $0.004 \pm 0.010$ when comparing the sample of galaxies projected
within $1\times R_{200}$ and the reference sample of galaxies projected at $[3-4]\times R_{200}$ away from the cluster
centers.

The relation between the total mass of intracluster dust, M$_d$, and average visual extinction, A$_V$, within a radius
R$_{clust}$ can be expressed as (\citealt{whi92}):
\begin{equation}
M_d = \frac{16 \pi}{9} \frac{A_V}{1.086}  \frac{r_g\rho_g}{Q_{ext}} R_{clust}^2,
\end{equation}
where the dust grains are characterized by their radius r$_g$ ($\sim 0.1 \mu$m), mass density $\rho_g$
($\sim 1$ g~cm$^{-3}$), and extinction efficiency factor Q$_{ext}$ ($\sim 2$). The values given in parenthesis are
characteristic for standard silicate grains (\citealt{hil83}). There is, however, no guarantee that ICM dust grains
have similar properties. Should there be dust with such properties and uniformly distributed within $R_{clust}
\sim 1.5$ Mpc (roughly corresponding in average to $1\times R_{200}$) from the cluster center, the $3\sigma$ upper
limit on A$_V$ would yield a dust mass of M$_d \sim 8 \times 10^9$ M$_\odot$.

\cite{che07} have recently claimed the detection of ICM dust by studying both the photometric and spectroscopic
properties of background quasars at different projected distances from galaxy clusters, using SDSS data. They
report reddening of few $10^{-3}$ magnitudes over large scale up to $[6-7]\times R_{200}$. The slight difference
with respect to our results could be attributed either to systematic errors or to the fact that they used
relatively closer clusters $0.1 < z < 0.3$. Despite a larger number of background sources in our data, our error
bars are about $2-3$ times larger. This could be explained by the fact that the color distribution of quasars is
narrower than that of galaxies.
 
Indeed, under Gaussian probability distribution, the uncertainty on the mean can
be expressed as $\sigma / \sqrt{N}$, where $\sigma$ is the dispersion of the distribution, and N, the number of
elements. The dispersion in color is $\sim 0.5$ for SDSS quasars (see, e.g., \citealt{ric01}), and $\sim 2$ for
our galaxies samples (Fig.\ref{color}). Therefore, with N = 8,000 background galaxies in our sample $a$, and
N = 3,000 background quasars in \cite{che07} study, we naturally end up with uncertainties $\sim 2.5 \times$
larger in our case. The one way to improve our results would then be to significantly increase the number of
background galaxies, for example with a larger survey. We note that, within a radius of $1 \times R_{200}$, we
get about 20 background galaxies per cluster, whereas \cite{che07} count a total of about 3,000 quasars for their
sample of $10^4$ clusters, i.e., in average, less than one background quasar per cluster. 

Given the uncertainties associated with those measurements, it is still difficult to reach a definite conclusion about
the presence of dust in the intracluster medium. All the studies based on the measurement of reddening of background
sources (cf. \citealt{mao95, nol03, che07}; and this work) agree with null or very small average reddening, in contrast
to previous counts of background clusters or quasars (Table \ref{prevsearch}), which were probably severely affected by
selection effects. At infrared or submillimeter wavelengths, on the other hand, only the combination of high angular
resolution and sensitivity might help to clarify the claimed detection by \cite{sti98} and \cite{mon05}. Still, the
task will not be eased by the large angular size of galaxy clusters, as, e.g., emission from diffuse and extended ICM
dust might be filtered out by interferometers.

%The associated intensity $I_\nu$ of the dust thermal emission can be calculated as:
%\begin{equation}
%I_\nu = \mathcal{N}_d B_\nu(T_d) \pi r_g^2 Q_{ext}
%\end{equation}
%where $\mathcal{N}_d$ is the dust column density and B$_\nu$(T$_d$) the Planck function for the average dust temperature T$_d$.
%With T$_d = 20$ K (\citealt{dwe90,yam05}), we derive an upper limit for the 100 $\mu$m intensity of $\sim 8~10^5$ Jy/s
%%towards the inner $\sim 1$ Mpc of a cluster,
%again assuming that the dust in uniformly distributed within $1 \times R_{200}$.
%This value is about one order of magnitude higher than that obtained by \cite{mon05} towards cluster center from the 
%co-addition of IRAS maps.

\section{Summary and Conclusions}

In this paper, we report a search for the presence of dust in the intracluster medium. Our method relies on the
statistical measurement of the color differences between galaxies located in the background of a cluster and
reference galaxies. We take advantage of data acquired with the CFHT in the frame of the Red Sequence Cluster Survey,
providing us with deep and homogeneous catalogs of galaxies and galaxy clusters. For our purpose, a major asset of
these RCS data is the availability of photometric redshifts for more than a million galaxies. We extracted 458
clusters up to a redshift of 0.5, and a total of $\sim 90,000$ galaxies, selected from their photometric redshift
$0.5 < z_{ph} < 0.8$ and with photometric redshift uncertainty $\delta z_{ph}/(1+z_{ph}) < 0.06$. In particular, our
selection criteria ensure low contamination by foreground galaxies and cluster members. Special attention was given
to the propagation of photometry and photometric redshift uncertainties, which were taken into account by a combination
of Monte-Carlo and bootstrap methods.

A sample of about 8,000 galaxies projected within $1\times R_{200}$ of a cluster is compared to different reference
samples composed of galaxies projected at distances [1-2], [2-3], [3-4] and [6-7]$\times R_{200}$ from a cluster center.
We do not detect significant color differences between the first sample and the different references. Unless intracluster
dust extends over scales larger than $\sim 10$ Mpc in radius, we therefore conclude that the dust extinction is
statistically low, with visual extinction $<$A$_V$$>$ = $0.004 \pm 0.010$ for the 458 RCS clusters with $z < 0.5$,
corresponding to an upper dust mass limit of $8 \times 10^9$ M$_\odot$ within $1 \times R_{200}$.
Given the uncertainties, our results are consistent with the recent measurements of reddening of background quasars
by galaxy clusters from \cite{che07}.

The on going RCS2 survey, extending the RCS coverage to $\sim 1,000$ deg$^2$ in four photometric bands with the CFHT
MegaCam will soon be available. The observed sky surface will be multiplied by a factor of 30, providing a much larger
number of background galaxies. These new data could allow us to improve our measurements and compare
the dust reddening towards clusters with different properties, such as richness, X-ray brightness or merging activity.

\acknowledgements
Based on observations obtained at the Canada-France-Hawaii Telescope (CFHT) which is operated by the
National Research Council of Canada, the Institut National des Sciences de l'Univers of the Centre
National de la Recherche Scientifique of France, and the University of Hawaii.

\begin{deluxetable}{llll}

%\rotate
\tabletypesize{\scriptsize}

\tablecaption{Previous searches for ICM dust.\label{prevsearch}}

%\label{prevsearch}

\startdata

\tableline
\tableline

\multicolumn{1}{c}{Method} & \multicolumn{1}{c}{Targets} & \multicolumn{1}{c}{Results} & \multicolumn{1}{c}{Reference} \\
\hline

& & & \\
\multicolumn{4}{l}{{\bf Infrared emission}} \\
& & & \\

Submm (JCMT)             & 11 clusters & no detection & \cite{ann93} \\

FIR (IRAS)               & 56 clusters & 2 clusters with some extended & \cite{wis93} \\
                         &             & FIR excess emission & \\

FIR (ISO)                & Coma  &  $0.01 < {\rm A}_V < 0.2$ & \cite{sti98} \\

%submm (JCMT)             & Central galaxies of 2 out  & & \cite{edg99} \\
%                         & of 5 cooling flow clusters & & \\

FIR (ISO)                & 5 Abell clusters & A$_V << 0.1$ & \cite{sti02} \\

Co-addition of IRAS data & 11507 clusters & statistical detection & \cite{mon05} \\ 

& & & \\
\multicolumn{4}{l}{{\bf Extinction and/or reddening of background sources}} \\
& & & \\

Background clusters      & Coma & A$_V \sim 0.4$ & \cite{zwi62} \\

Background clusters      & Coma & A$_V \sim 0.3$ & \cite{kar69} \\
                         & 15 clusters  & A$_V \sim 0.2 \pm 0.05$ & \\

Background clusters      & Abell clusters & A$_V \sim 0.3$ & \cite{bog73} \\

Background QSR           & Rich Abell clusters & A$_B \sim 0.5$ & \cite{rom92} \\

Background UVX objects  &  & A$_B = 0.2$ & \cite{boy88} \\

Background QSR           & Abell clusters &  E($B-V$) $< 0.05$ & \cite{mao95} \\

Background SDSS QSR      & $\sim 10^4$ SDSS clusters & E(g-i) = $0.003 \pm 0.002$ (photometry) & \cite{che07} \\
                         &                           & E(g-i) = $0.008 \pm 0.003$ (spectroscopy) & \\

Background galaxies      & 140 APM clusters & A$_R$ and E($B-R) \leq 0.025$ & \cite{nol03} \\

Background galaxies      & 458 RCS clusters & $<$E($B-R_c$)$>$ = $0.005 \pm 0.008$  & {\bf this work} \\
                         &                  & $<$E($V-z'$)$>$ = $0.000 \pm 0.008$   & \\

& & & \\
\multicolumn{4}{l}{{\bf Others methods}} \\
& & & \\

Redshift asymmetry & nearby galaxy groups & A$_B \sim 0.1 - 0.2$  & \cite{gir92} \\
\hspace{0.3cm} and color-velocity correlation & & & \\

Ly$\alpha$/H$\alpha$ ratio & 10 cooling flow clusters& E($B-V$) $\sim 0.19$ & \cite{hu92} \\

Mg$_2$ {\em vs} $B-V$ correlation & 19 nearby clusters & E($B-V$) $< 0.06$ & \cite{fer93} \\

\enddata
\end{deluxetable}

\begin{table}
\begin{center}
\caption{Subset of used RCS survey data.}
\label{RCSsubset}

\begin{tabular}{cccccc}

\hline
\hline

Patch & R.A.    & Dec.    & Area \tablenotemark{a} & Number of     & Number of \\
name  & (J2000) & (J2000) & (deg$^2$) & selected      & selected     \\
      &         &         &           & clusters \tablenotemark{b} & galaxies \tablenotemark{c} \\

\hline

RCS0226 & 02 26 07.0 & +00 40 35   & \phantom{\phn}3.99 & \phantom{\phn}64 & 11,119  \\
RCS0351 & 03 51 20.7 & $-$09 57 41 & \phantom{\phn}4.30 & \phantom{\phn}52 & \phantom{\phn}1,679  \\
RCS0926 & 09 26 09.6 & +37 10 12   & \phantom{\phn}4.85 & \phantom{\phn}70 & 18,731  \\
RCS1122 & 11 22 22.5 & +25 05 55   & \phantom{\phn}4.72 & \phantom{\phn}74 & \phantom{\phn}9,154  \\
RCS1328 & 13 27 41.9 & +29 43 55   & \phantom{\phn}1.34 & \phantom{\phn}24 & \phantom{\phn}1,023  \\
RCS1416 & 14 16 35.0 & +53 02 26   & \phantom{\phn}3.04 & \phantom{\phn}40 & 12,056  \\
RCS1449 & 14 49 26.7 & +09 00 27   & \phantom{\phn}2.01 & \phantom{\phn}21 & \phantom{\phn}5,913  \\
RCS1616 & 16 16 35.5 & +30 21 02   & \phantom{\phn}4.16 & \phantom{\phn}48 & 14,142  \\
RCS2153 & 21 53 10.8 & $-$05 41 11 & \phantom{\phn}2.96 & \phantom{\phn}38 & \phantom{\phn}9,573  \\
RCS2318 & 23 18 10.7 & $-$00 04 55 & \phantom{\phn}2.23 & \phantom{\phn}27 & \phantom{\phn}6,940  \\
\hline
Total   &            &             & 33.60 & 458 & 90,330  \\

\hline

\tablenotetext{a}{Area with data in $B, V, R_c$ and $z'$ filters.}
\tablenotetext{b}{Clusters with $z_{clust}$ $< 0.5$ and Bgc $> 200$.}
\tablenotetext{c}{Galaxies with $0.5 < z_{ph} < 0.8$  and $\delta z_{ph}/(1+z_{ph}) < 0.06$}

\end{tabular}
\end{center}
\end{table}

\begin{table}
\begin{center}
\caption{Results.} \label{results}
\label{percent}
\begin{tabular}{cccc}
\hline
\hline
Samples & $<$E($B-R_c$)$>$ & $<$E($V-z'$)$>$ & $<$A$_V$$>$ \tablenotemark{a} \\
 & \multicolumn{3}{c}{(10$^{-3}$ mag.)} \\
\hline

$[0-1]$ {\em vs} $[1-2]\times R_{200}$ & $1 \pm 8$ & $-1 \pm 8$ & \phantom{\phs}$0 \pm 11$ \\
$[0-1]$ {\em vs} $[2-3]\times R_{200}$ & $5 \pm 8$ & $-2 \pm 8$ & \phantom{\phs}$3 \pm 11$ \\
$[0-1]$ {\em vs} $[3-4]\times R_{200}$ & $5 \pm 8$ & \phantom{\phs}$0 \pm 8$ & \phantom{\phs}$4 \pm 10$ \\
$[0-1]$ {\em vs} $[6-7]\times R_{200}$ & $5 \pm 7$ & $-4 \pm 7$ & \phantom{\phs}$1 \pm 9$\phantom{\phn} \\

$[1-2]$ {\em vs} $[2-3]\times R_{200}$ & $4 \pm 5$ & $-1 \pm 4$ & \phantom{\phs}$2 \pm 6$\phantom{\phn} \\

$[1-2]$ {\em vs} $[6-7]\times R_{200}$ & $3 \pm 4$ & $-2 \pm 4$ & \phantom{\phs}$1 \pm 5$\phantom{\phn} \\
$[2-3]$ {\em vs} $[6-7]\times R_{200}$ & $0 \pm 3$ & $-2 \pm 3$ & $-2 \pm 3$\phantom{\phn} \\
$[3-4]$ {\em vs} $[6-7]\times R_{200}$ & $0 \pm 2$ & $-2 \pm 2$ & $-2 \pm 3$\phantom{\phn} \\

\hline
\end{tabular}

\tablenotetext{a}{The visual extinction, $<$A$_V$$>$, is calculated from a weighted average of the
two independent reddening values of the previous columns, assuming the relations
A$_V = 1.9 \times$E($B-R_c$), and A$_V = 1.9 \times$E($V-z'$) [from \cite{sch98}].}
\end{center}
\end{table}

\end{document}